\documentclass[twocolumn,showpacs,preprintnumbers,amsmath,amssymb]{revtex4}

\usepackage{graphicx}
\usepackage{dcolumn}

\usepackage{amsmath,amssymb}

\usepackage[mathscr]{euscript}

\usepackage{calc}

\usepackage{nicefrac}

\usepackage{xcolor}

\usepackage{mathptmx, courier, pifont}
\usepackage[scaled=0.92]{helvet}
\usepackage[T1]{fontenc}
\usepackage{textcomp} 

\usepackage{bm}

\newcommand{\eq}[1]{Eq.\ (\ref{#1})}
\newcommand{\eqnoeq}[1]{(\ref{#1})}
\newcommand{\transpose}{^{\rm\scriptscriptstyle T}}
\newcommand{\taucrit}{\tau_{\rm c}}

\newcommand{\ndensity}{\mathscr{N}}
\newcommand{\phaseden}{\mathscr{F}}

\newcommand{\rin}{{\mathscr{R}^{\rm in}}}
\newcommand{\rincomponent}[2]{{\mathscr{R}^{\rm in}_{#1#2}}}
\newcommand{\rinhatcomponent}[2]{{\hat{\mathscr{R}}^{\rm in}_{#1#2}}}
\newcommand{\rout}{{\mathscr{R}^{\rm out}}}
\newcommand{\routcomponent}[2]{{\mathscr{R}^{\rm out}_{#1#2}}}
\newcommand{\routhatcomponent}[2]{{\hat{\mathscr{R}}^{\rm out}_{#1#2}}}
\newcommand{\rinout}{{\mathscr {R}^{\rm in/out}}}
\newcommand{\rhatinout}{\hat{\mathscr{R}}^{\rm in/out}}
\newcommand{\rinoutcomponent}[2]{\mathscr{R}^{\rm in/out}_{#1#2}}

\newcommand{\units}[1]{\mbox{\,#1}}
\newcommand{\punits}[2]{\units{#1}^{#2}}
\newcommand{\script}[1]{\mathscr{#1}}
\newcommand{\error}{\script E}

\newcommand{\singlefig}[6]{%
\begin{figure}[tpb]\vspace{#3}%
\includegraphics*[scale=#5]{#2}%
\caption{\label{fig:#1} #6}%
\vspace{#4}%
\end{figure}}

\newcommand{\doublefig}[6]{%
\begin{figure*}[tpb] \vspace{#3}%
\includegraphics*[scale=#5]{#2}%
\caption{\label{fig:#1} #6}
\vspace{#4}
\end{figure*}}
\newcommand{\runningheads}[2]{\markboth{\hfill #1\hfill}{\hfill #2\hfill}}

\begin{document}

\title{Fast Explicit Solutions for Neutrino-Electron Scattering:  Explicit  Asymptotic Methods}

\author{Aaron Lackey--Stewart$^{(1)}$}
\email{alackeyste@hotmail.com}
\author{Raghav Chari$^{(1)}$}
\email{rchari1@vols.utk.edu}
\author{Adam Cole$^{(1)}$}
\email{acole32@vols.utk.edu}
\author{Nick Brey$^1$}
\email{nbrey@vols.utk.edu}
\author{Kyle Gregory$^{(1)}$}
\email{kyle_gregory_2012@yahoo.com}
\author{Ryan Crowley$^{(1)}$}
\email{jcrowle7@vols.utk.edu}
\author{Mike Guidry$^{(1)}$}
\email{guidry@utk.edu}
\author{Eirik Endeve$^{(1,2)}$}
\email{endevee@ornl.gov}

\affiliation{
$^{(1)}$Department of Physics and Astronomy, University of Tennessee,
Knoxville, TN 37996-1200, USA
\\
$^{(2)}$Computer Science and Mathematics Division, Oak Ridge National
Laboratory, Oak Ridge, TN 37830, USA
}

\date{\today}

\begin{abstract}

We present results of explicit asymptotic approximations applied to 
neutrino--electron scattering in a representative model of neutrino population evolution 
under conditions characteristic of core-collapse supernova explosions or binary neutron star mergers. It is shown that this approach provides stable solutions of these 
stiff systems of equations, with accuracy and timestepping comparable to that 
for standard implicit treatments such as backward Euler, fixed point iteration, 
and Anderson-accelerated fixed point iteration. Because each  timestep can be 
computed more rapidly with the explicit asymptotic approximation than with 
implicit methods, this suggests that algebraically stabilized explicit 
integration methods could be used to compute neutrino evolution coupled to hydrodynamics more efficiently in stellar explosions and mergers than the methods currently in use.

\end{abstract}


\maketitle

\runningheads{}{\em Fast Explicit Solutions for Neutrino-Electron Scattering: Explicit Asymptotic Methods}


\section{\label{h:intro} Introduction}

Computer simulations of events like core collapse supernova explosions or 
neutron star mergers require solving simultaneously coupled systems of non-linear partial differential equations governing hydrodynamics, neutrino radiation transport, and thermonuclear reactions. Because rate parameters entering into these processes can differ by many orders of magnitude, such systems are computationally \textit{stiff} and stability demands that great care be taken in choosing methods to solve them~\cite{oran05, gear71, lam91, press92}.  Explicit methods are generally easier to implement than implicit ones because  they require only the current state of the system to compute a future state, whereas implicit methods require also information at the next step, which is initially unknown so an iterative procedure is necessary. 
This iteration often involves matrix inversions at each step that are increasingly 
expensive for larger sets of equations.
Thus  computing an explicit timestep is generally faster and less memory  intensive than computing an implicit timestep for large networks because they do not require matrix inversions.  However  traditional explicit  methods such as forward Euler are impractical for solving the extremely stiff systems  characteristic of many astrophysical processes because maintaining stability restricts the integration timesteps to tiny values, and this far  outweighs the faster explicit computation of each timestep.  

Thus standard explicit  methods have usually been thought impractical for stiff systems,  forcing the use of stable but more complex implicit methods.  Because of the matrix inversions, these  implicit methods are often prohibitively slow for networks of realistic size and  complexity when coupled to a large-scale fluid dynamics simulation.  This has  typically forced the use of drastic approximations, often employing implicit networks that are stable but too small and/or too schematic to be realistic.  We characterize such approximations as {\em  poorly controlled}, as they introduce systematic errors that can be estimated only qualitatively because they are dictated by the ability to integrate the network with the chosen method rather than by the realistic physics of the network.  But in a large-scale  simulation of say a supernova, typically a  multidimensional hydrodynamical code is coupled to kinetic networks (coupled  ordinary differential equations) representing thermonuclear burning and/or  neutrino evolution, and it may be argued that these kinetic networks need be  solved only {\em approximately,} provided that the approximation introduces  {\em controlled errors} that are  smaller than the errors and  uncertainties associated with the overall coupled hydrodynamical--kinetic  system. 

Given the current state of the art in such simulations, a conservative  rule of thumb is that (controlled) errors of a few percent or less are acceptable for  approximate kinetic networks, since to do better wastes resources and will not improve the final  results. This raises the question of  whether there exist approximations for standard explicit methods that could allow  taking more competitive timesteps with an acceptable controlled error, thus  permitting the solution of larger and more realistic kinetic networks coupled to fluid dynamics.

In a series of publications  \cite{guidJCP,guidAsy,guidQSS,guidPE,fege2011,chup2008,brey2022,broc2015,  haid2016,guid2023a,brey2023}, our group has  demonstrated a new class of {\em algebraically stabilized explicit approximations} for thermonuclear networks that extends earlier work by Mott and  collaborators \cite{mott00,mott99}.  These new explicit methods use algebraic  constraints to stabilize explicit integration steps, resulting in stable  timesteps that are comparable to those of standard  implicit methods, even for extremely stiff systems of equations where  the fastest  and slowest rates can differ by 10-20 orders of magnitude. These methods require matrix--vector multiplication but not matrix inversions, so they scale more gently with network size than  implicit algorithms. This more favorable scaling coupled with a competitive  integration stepsize allowed a demonstration that such algorithms are capable of  performing numerical solution of extremely stiff kinetic networks containing  hundreds of species 5-10 times faster than standard implicit methods on a single  CPU (Central Processing Unit) \cite{guidJCP,guidAsy,guidQSS,guidPE}. Furthermore,  we found these explicit algebraic methods to be extremely amenable to GPU (Graphical  Processing Unit) acceleration, permitting of order $1000$ large (150-isotope) thermonuclear networks to be executed in parallel on a single GPU in the time required for one such network to be solved by standard implicit means \cite{broc2015,haid2016}.

The demonstrations described above were for astrophysical thermonuclear networks  but there is no fundamental reason why such methods cannot be applied to other  stiff kinetic networks in many scientific applications, since most such networks  have a  mathematical and computational structure similar to that of thermonuclear networks \cite{guidJCP}.  To this end, we note that an adequate description of neutrino transport often is of fundamental importance for very high-temperature, very high density environments like those found in core collapse supernovae or binary neutron star mergers. Most large-scale simulations dealing with neutrinos have been forced to make significant approximations in the neutrino transport, and often in such simulations the largest fraction of the computing budget is allocated to evolution of neutrinos, even when they are treated approximately \cite{mezzacappa_etal_2020,foucart_2023}.  Accordingly, we have begun developing codes that apply explicit algebraic algorithms similar to those developed for thermonuclear networks to the evolution of neutrino populations.  

Our suite of neutrino codes is being written primarily in C++ and will be referred to collectively as FENN (\underline{F}ast \underline{E}xplicit \underline{N}eutrino \underline{N}etwork).  It is our intention to make FENN available as a documented, open source, community resource. In this  paper we present the first application of the new explicit algebraic methods implemented in FENN to  evolution of neutrino populations due to neutrino-electron scattering (NES) under the hot, dense conditions characteristic of explosions and mergers.  
Mezzacappa \& Bruenn \cite{mezzacappaBruenn_1993c} demonstrated the impact of a realistic treatment of NES, as opposed to more approximate Fokker--Plank treatments \cite{myra_etal_1987}, on the dynamics of stellar core collapse.  
Down-scattering mediated by NES increases the neutrino mean free path, which can result in increased deleptonization, and a reduced electron fraction in the collapsed core.  
See also \cite{lentz_etal_2012a,lentz_etal_2012b,just_etal_2018} for further discussions on the role of NES and the complex interplay of neutrino opacities in supernova models.  
It is generally accepted that NES is part of the minimal set of opacities to include in realistic models \cite{burrows_etal_2006,janka_2012,fischer_etal_2017,mezzacappa_etal_2020}.  

Since NES couples neutrinos across momentum space, their inclusion in realistic models is computationally taxing and approximations have been employed to lessen the burden, including artificially reducing the scattering kernel to allow for explicit integration \cite{thompson_etal_2003,oConnor_2015,just_etal_2018}.  
Here we investigate the utility of approximating the NES operator to enable explicit integration with controlled errors.  
We shall find strong evidence that speedups relative to standard implicit methods similar to those found for thermonuclear networks hold also for neutrino networks, suggesting a viable  path to faster solution of neutrino transport equations through algebraically  stabilized explicit integration.

\section{Mathematical model of Neutrino-Electron Scattering\label{mathmod}}

The basic problem is formulated as the evolution of a spectral neutrino number 
distribution discretized as a set of energy bins, each of phase-space 
volume $V$, with neutrino scattering into and out of the energy bins governed by 
the Boltzmann equation. For this particular model we use a Boltzmann equation 
that is spatially homogeneous, with inelastic neutrino--electron scattering from 
a  matter background that remains constant during scattering, implying that~\cite{bru85},
\begin{align}
     \frac{d\phaseden}{dt} &= \left(1 -
    \phaseden\right)\int_{V_{P}}\rin \left( \epsilon , \epsilon', 
    \bm n \cdot \bm n' ; \bm u \right)\phaseden' dV_{P'}
    \nonumber
    \\
    &\quad- \phaseden \int_{V_{P}}\rout\left( \epsilon , \epsilon', 
    \bm n \cdot \bm n ' ; \bm u \right)\left(1 - \phaseden' 
    \right)dV_{P'},
    \label{boltzmann}
\end{align}
where $\phaseden = \phaseden(\epsilon, \omega, t)$ is the phase-space  density normalized to lie in the interval $[0,1]$, the neutrino energy is  $\epsilon$, the neutrino propagation direction is specified by $\omega =  \omega(\vartheta, \phi)$ with $\vartheta \in [0, \pi]$ and $\phi \in [0, 2\pi]$,  the time is $t$, we define a unit 3-vector $\bm n = (\cos \vartheta,\ \sin  \vartheta \cos \phi,\ \sin\vartheta \sin \phi)\transpose$  (with {\small T}  denoting the matrix transpose). The kernels  $\rin(\epsilon , \epsilon', \bm n \cdot \bm n ' ; \bm u)$ and $\rout(\epsilon , \epsilon', \bm n \cdot \bm n ' ; \bm u)$  appearing in \eq{boltzmann} represent transition rates into or out of a  bin in momentum space, respectively, which are functions of neutrino energy before and after  collision and the cosine of the angle $\alpha$ between the unit three-vectors  $\bm n$ and $\bm n'$. The thermodynamic state of the ambient  matter is given by $\bm u = (\rho, T, Y_{e})\transpose$, where $\rho$ is the  mass density, $T$ is the temperature, and $Y_{e}$ is the electron fraction.  The blocking factors $(1 - \phaseden)$ and $(1 - \phaseden')$ vanish when the  respective phase space volume is full, so that neutrino transitions are  prohibited  and the Pauli exclusion principle is obeyed. Finally, the  momentum-space volume element is $dV_{P} = dV_{\epsilon}d\omega$, where  $dV_{\epsilon} = \epsilon^{2} d\epsilon$ is the volume element of an energy  shell.

It will be most convenient to express \eq{boltzmann} in terms of the spectral number 
density of neutrinos $\ndensity$. To simplify the current analysis it will be assumed that scattering is isotropic during neutrino propagation. To implement this we first express 
the scattering kernel as the expansion~\cite{smit96},
\begin{equation}
    \rinout(\epsilon, \epsilon', \cos \alpha, 
    \bm u) \approx \sum_{\ell = 0}^{L} 
    \Phi_{\ell}^{\rm in/out}(\epsilon, \epsilon', 
    \bm u)P_{\ell}(\cos \alpha),
    \label{legendre-expansion}
\end{equation}
where $P_{\ell}(\cos \alpha)$ is the $\ell^{th}$ Legendre polynomial. From the 
orthogonality of the Legendre polynomials,
\begin{multline}
    \Phi_{\ell}^{\rm in/out}(\epsilon, \epsilon', \bm u) = 
    \\
    \frac{2\ell + 1}{2} \int_{-1}^{1} \rinout(\epsilon, \epsilon', 
    \cos\alpha, \bm u)P_{\ell}(\cos \alpha)d\cos \alpha .
\end{multline}
Then the scattering may be approximated as isotropic by considering only the first 
term in the Legendre expansion given in \eq{legendre-expansion}, which removes 
the angular dependence from the scattering kernels. Lastly, we  integrate away 
the $\omega(\vartheta,\phi)$ dependence using the relation~\cite{thor81}: 
\begin{equation}
    \ndensity(t) = 
    \frac{1}{4\pi}\int_{0}^{2\pi} \int_{0}^{\pi} \phaseden(\epsilon, \textbf{x}, t) \sin \vartheta 
    d\vartheta d\phi.
    \label{thorne}
\end{equation}
This approach uses a zeroth-moment formalism simulation for isotropic neutrino scattering. Using \eq{thorne} and substituting the first  term of the Legendre expansion \eqnoeq{legendre-expansion}, we set  $\rinout = \Phi_{0}^{\rm in/out}$ and the number density  representation of \eq{boltzmann} is given by 
\begin{align}
    \frac{d\ndensity}{dt}(\epsilon)&= (1 - \ndensity(\epsilon))\int_{\mathbb{R}^{+}} 
    \rin(\epsilon, \epsilon')\ndensity (\epsilon')dV_{\epsilon'}
    \nonumber
    \\
    &\quad- 
    \ndensity(\epsilon)\int_{\mathbb{R}^{+}}\rout(\epsilon, \epsilon')(1 - 
    \ndensity(\epsilon'))dV_{\epsilon'} ,
    \label{dNdt}
\end{align}
where $\phaseden$ and $\ndensity$ take values between 0 and 1, a factor  of $4\pi$ has been absorbed into $dV_{\epsilon'}$, and  the  integrals may be interpreted as being over the surface area of the energy  shell. Note that the number of particles must be conserved during the  scattering process~\cite{cern94},
\begin{equation}
    \rin(\epsilon, \epsilon') = 
    \rout(\epsilon', \epsilon) ,
    \label{symmScattKern}
\end{equation}
and that for inverse temperature $\beta \equiv T^{-1}$, the equilibrium condition, 
\begin{equation}
    \rin(\epsilon, \epsilon') = 
    \rout(\epsilon, \epsilon')e^{\beta(\epsilon' - \epsilon)},
\end{equation}
 must be satisfied.
Then when the exchange in energy and momentum  induced by matter--neutrino scattering disappears, $\ndensity$ tends to the  equilibrium Fermi--Dirac  distribution,
\begin{align}
    \ndensity_{\rm eq}(\epsilon, \bm u) &= 
    \frac{1}{4\pi}\int_{0}^{2\pi} \int_{0}^{\pi} \phaseden_{\rm eq} \sin 
    \vartheta 
    d\vartheta d\phi 
    \nonumber
    \\
    &= \frac{1}{e^{\beta(\epsilon - \mu_{\nu})} + 1} ,
\end{align}
where $\phaseden_{\rm eq}$ is the phase space density at equilibrium and 
$\mu_\nu$ is the neutrino chemical potential.

\section{Discretization of the model\label{discretization}}

For computational purposes let us now discretize the energy domain into $N_b$ energy bins, $ 0 = \epsilon_{\frac{1}{2}}<\epsilon_{\frac{3}{2}}<...<\epsilon_{N_b + \frac{1}{2}} = \epsilon_{Max}$, with each bin  having a center $\epsilon_{i}$  given by
\begin{equation}
    \epsilon_{i} = \frac{\epsilon_{i - 1/2} + \epsilon_{i -1/2}}{2},
    \label{bin-center}
\end{equation}
and a volume $\Delta V_{i}^{\epsilon}$  given by
\begin{equation}
    \Delta V_{i}^{\epsilon} = \int_{\epsilon_{i - 
    1/2}}^{\epsilon_{i + 1/2}} dV_{\epsilon} = \frac{4\pi}{3}\left(\epsilon_{i + 
    1/2}^{3} 
    - \epsilon_{i - 1/2}^{3}\right).
    \label{bin-volume}
\end{equation}
We then discretize \eq{dNdt} as
\begin{align}
    \frac{d\ndensity}{dt} &\approx (1 - \ndensity)\int_{D^{\epsilon}} 
    \rin(\epsilon, \epsilon')\ndensity (\epsilon')dV_{\epsilon'} 
    \nonumber
    \\
    &\quad- 
    \ndensity\int_{D^{\epsilon}}\rout(\epsilon, \epsilon')\big(1 
    -\ndensity(\epsilon')\big)dV_{\epsilon'},
    \label{discretedNdt}
\end{align}
where the integrals are now over a finite energy domain $D^{\epsilon} = [0, \epsilon_{Max}]$.  Because of the neutrino--matter interactions  neutrinos will scatter into and out of each bin, so the total $d\ndensity_{i}/dt$ for bin $i$ involves a sum of terms like \eq{discretedNdt}  over all bins, giving
\begin{align}
    \frac{d\ndensity_{i}}{dt} &= \sum_{k = 1}^{N_{b}}(1 - 
    \ndensity_{i})\int_{\epsilon_{k - 1/2}}^{\epsilon_{k + 1/2}} 
    \rincomponent ik (\epsilon, 
    \epsilon')\ndensity_{k}(\epsilon')dV_{\epsilon'} 
    \nonumber
    \\
    &\quad- \sum_{k = 
    1}^{N_{b}}\ndensity_{i}\int_{\epsilon_{k - 1/2}}^{\epsilon_{k + 
    1/2}}\routcomponent ik (\epsilon, \epsilon')\big(1 - 
    \ndensity_{k}(\epsilon')\big)dV_{\epsilon'},
    \label{dNdt2}
\end{align}
where the scattering kernels  $\rin$ and $\rout$ mediate the transitions into and out of each bin, respectively, with indices $ik$ representing scattering between the  $i^{th}$ bin with density $\ndensity_{i}$ and the $k^{th}$ bin with density $\ndensity_{k}$. Assuming constant scattering rates in each bin, the  kernels may be taken outside of the integrals in \eq{dNdt2}. Using  \eq{bin-volume} for the volume of each bin,   and the volume-averaged particle density in each   energy  bin, 
\begin{equation}
    \ndensity_{i}(t) = \frac{1}{\Delta 
    V_{i}^{\epsilon}}\int_{\epsilon_{i - 1/2}}^{\epsilon_{i + 1/2}} 
    \ndensity (\epsilon, t) dV_{\epsilon},
    \label{eq13}
\end{equation}
we  arrive at 
\begin{align}
    \frac{d\ndensity_{i}}{dt} = (1 - \ndensity_{i}) \sum_{k = 
    1}^{N_{b}} \rinhatcomponent ik \ndensity_{k} 
    - 
    \ndensity_{i}\sum_{k = 1}^{N_{b}}\routhatcomponent ik(1 - 
    \ndensity_{k}),
    \label{dN/dt3}
\end{align}
where $i = 1,2, \ldots , N_b$ and $\rhatinout = \rinout \Delta V_k^\epsilon$. 
Equation \eqnoeq{dN/dt3} describes the rate at which the  neutrino number density in the $i^{th}$ energy bin is changing,  with the  blocking factors $(1-\ndensity_i)$ and $(1-\ndensity_k)$ insuring that  the Pauli exclusion principle is obeyed. Further manipulation of \eq{dN/dt3} yields
\begin{equation*}
    \frac{d\ndensity_{i}}{dt} = 
    \sum_{k=1}^{N_{b}}\rinhatcomponent ik \ndensity_{k} - 
    \ndensity_{i} \sum_{k =1}^{N_{b}}\left[ \routhatcomponent ik + 
    (\rinhatcomponent ik - 
    \routhatcomponent ik )\ndensity_{k} , \right],
\end{equation*}
and upon defining 
\begin{equation}
\begin{gathered}
    \eta_{i} \equiv 
    \sum_{k=1}^{N_{b}}\rinhatcomponent ik \ndensity_{k}
    \qquad
    \kappa_{i}\equiv 
    \sum_{k =1}^{N_{b}}\routhatcomponent ik
    \\
    \tilde{\kappa}_{i} \equiv  \sum_{k =1}^{N_{b}}\left [ 
    \delta_{ik}\kappa_{k} + (\rinhatcomponent ik - 
    \routhatcomponent ik)\ndensity_{k} \right ],
    \label{definitions1}
    \end{gathered}
\end{equation}
the rate of change for the number density in bin $i$ may be written in the 
compact form 
\begin{equation}
    \frac{d\ndensity_{i}}{dt} = \eta_{i} - 
    \tilde{\kappa}_{i}\ndensity_{i} = C_{i},
    \label{dNdt4}
\end{equation}
where 
$
    \eta_i = \sum_{k}\rinhatcomponent ik \ndensity_{k}
$ 
represents the total flux of neutrinos into the $i^{th}$ bin  and $\tilde{\kappa}_{i}\ndensity_{i}$ represents a flux depleting the  $i^{th}$ bin due to neutrino scattering with a rate $\tilde{\kappa}_{i}$. If $\eta_{i}$ and   $\tilde{\kappa}_{i}$ are held constant, \eq{dNdt4} has an exact solution
 \begin{equation}
     \ndensity_{i}(t) = \ndensity_{0}e^{- \tilde{\kappa}_{i}t} 
    + \frac{\eta_{i}}{\tilde{\kappa}_{i}}\big(1 - e^{- \tilde{\kappa}_{i}t}\big),
 \end{equation}
where $\ndensity_{0}$ is the initial neutrino number density. Thus 
$\tilde\tau_i = \tilde{\kappa}_{i}^{-1}$ may be interpreted as the timescale over which  
bin $i$ reaches a steady state, and a mean free path $ \ell_i$ for neutrinos in 
energy/momentum bin $i$  may be defined as,
 \begin{equation}
     \ell_i = \frac{c}{\tilde{\kappa}_{i}},
     \label{meanfree}
 \end{equation}
 by assuming  that  the almost-massless neutrinos propagate at lightspeed $c$.

 \section{Matrix Formulation\label{matrixform}} 

Introducing a vector $\pmb{\ndensity} = (\ndensity_{1},  \ndensity_{2},\dots, \ndensity_{N_{b}})\transpose$, and an 
$N_b \times N_b$ collision matrix $\bm{M}$ with elements $M_{ik}$, \eq{dNdt4} may be  expressed as
\begin{equation}
    \frac{d\pmb{\ndensity}}{dt} = 
    \pmb{M}(\pmb{\ndensity})\,\pmb{\ndensity} \equiv \pmb{C}(\pmb{\ndensity})
    \qquad
    {M_{ik} \equiv \rinhatcomponent ik - \delta_{ik} 
    \tilde{\kappa}_{k}} .
    \label{dNdt-vec}
\end{equation}
Defining $\dot{\ndensity}_{i} \equiv  d\ndensity_{i} / dt$, we have explicitly
\begin{equation}
    \begin{pmatrix}
    \dot{\ndensity}_{1} \\ \dot{\ndensity}_{2} \\  \dot{\ndensity}_{3} \\ 
    \vdots \\ \dot{\ndensity}_{N_{b}} \end{pmatrix}
    =
    \begin{pmatrix}
    M_{11} & M_{12} & M_{13} & \dots & M_{1N_{b}} \\
    M_{21} & M_{22} & M_{23} & \dots & M_{2N_{b}} \\
    M_{31} & M_{32} & M_{33} & \dots & M_{3N_{b} } \\
    \vdots   & \vdots   & \vdots   & \dots & \vdots \\ 
    M_{N_{b}1} &  M_{N_{b}2}  & M_{N_{b}3} & \dots & M_{N_{b}N_{b}}
    \end{pmatrix}
    \begin{pmatrix}
    \ndensity_{1} \\ \ndensity_{2} \\ \ndensity_{3} \\ \vdots \\ 
    \ndensity_{N_{b}} \end{pmatrix} 
    \label{dNdt-matrix}
\end{equation}
Thus, evolution of the neutrino distribution requires  integration of the matrix differential equation \eqnoeq{dNdt-matrix} with  appropriate initial conditions.  For explicit methods this requires relatively  inexpensive matrix--vector multiplications at each time step, while implicit  methods require also much more time-consuming matrix inversions at each timestep. An implicit algorithm typically must iterate multiple times at each timestep during periods of strong interaction, implying multiple matrix inversions per timestep, though implicit implementations can use Gauss--Seidel and other convergent matrix solvers to partially ameliorate the impact of these inversions. General experience with implicit  methods applied to sets of kinetic equations indicates that because of the  required matrix inversions the computing time scales from   quadratically to cubically with network size, making them very expensive for larger networks.

 \section{Stiffness\label{stiffstuff}}

The effective rate parameters entering the matrix in \eq{dNdt-matrix} 
depend on the number densities and  individual rates, and  often will differ among themselves by orders of magnitude. Thus the neutrino rate equations are stiff.   To illustrate the stiffness we consider the low occupancy 
limit $\ndensity_{k} \ll 1$ for $\tilde\kappa_i$ in \eq{definitions1}. Figure 
\ref{fig:meanCollision+meanFree}(a) shows a plot of the mean collision time 
$\tilde{\kappa}^{-1}$ as a function of energy for some spherically symmetric 
supernova models listed in Table \ref{tb:neutrinoModels}.
\doublefig
    {meanCollision+meanFree}
    {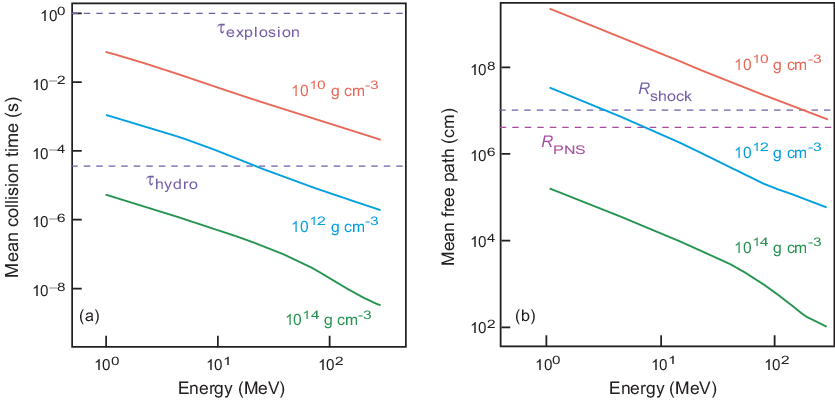}
    {0pt}
    {0pt}
    {1.00}
    {(a)~Mean neutrino collision times $\tilde{\kappa}^{-1}$ for three mass 
    densities (solid curves) corresponding to supernova models given in Table 
    \ref{tb:neutrinoModels}. Dashed horizontal lines indicate the characteristic explosion 
    time $\tau_{\rm explosion}$ and hydrodynamical timescale $\tau_{\rm hydro}$ for 
    a typical core-collapse supernova simulation. (b)~Neutrino mean free path for 
    three mass densities (solid curves). Dashed horizontal lines indicate a characteristic 
    shock radius $R_{\rm shock}$ and radius of the protoneutron star $R_{\rm PNS}$ 
    in a typical core-collapse supernova simulation.
}
For comparison, we display also a characteristic supernova explosion time and a  characteristic hydrodynamical timescale in  Fig.\ \ref{fig:meanCollision+meanFree}(a). For the highest density case of $10^{14}\punits{g\,cm}{3}$ in Fig.\ \ref{fig:meanCollision+meanFree}(a), we see  that the mean collision time is $ \sim10^{-5}$ seconds for the lowest energies and $  \sim10^{-9}$ seconds for the highest energy. Figure  \ref{fig:meanCollision+meanFree}(a) indicates that a standard  explicit integration is impractical for this problem.  The maximum stable  explicit timestep will be governed by the fastest timescale for the neutrino  interactions, which is about $10^{-9}$ seconds for the $10^{14}\punits{g\,cm}{3}$ case, but the timescale of the explosion is $\sim 1$  second.  Thus a traditional explicit integration method such as the forward Euler  algorithm  would require billions of  timesteps to integrate over the complete  explosion because of the stability constraint, whereas a typical  implicit calculation may require only of order hundreds of integration  steps for this case, because the implicit timestep typically is limited by accuracy but not by  stability.

\begin{table}
    \centering
    \caption{Conditions from supernova models \cite{lieb2004}\label{tb:neutrinoModels}} 
    \setlength{\tabcolsep}{5 pt}
    \begin{tabular}{cccc}
    \hline
    Model & Density $({\rm g\,cm}^{-3})$ & $kT$ (MeV) & $Y_{\rm e}$ \\
    \hline
    I   &    $1.0 \times 10^{14}  $ & 20.54  & 0.25   \\
    II  &    $1.0 \times 10^{12}  $ & 7.71   & 0.12   \\
    III &    $1.0 \times 10^{10}  $ & 3.14   & 0.26   \\
    \hline
    \end{tabular}
\end{table}

 \section{Neutrino Transport Regimes \label{transportRegimes}}
 
 Figure \ref{fig:meanCollision+meanFree}(b) shows the neutrino mean free path as a function of their energy for the models in Table \ref{tb:neutrinoModels}. As  expected intuitively based on the characteristic scaling of neutrino cross sections with  energy  and density, collisions are more common (the mean free path is shorter) at higher densities and  higher energies. Estimates of the size of the protoneutron star and the  supernova shock  radius calculated by \citet{bru2016} are also displayed, which shows that neutrinos  are trapped at the higher densities and are able to escape the protoneutron  star and the shock radius for low-energy neutrinos at low densities. Thus the  models that we  shall calculate span the range of neutrino transport behavior  from diffusion to transitional to free-streaming under core-collapse supernova conditions.

\section{Explicit Algebraic Methods\label{explicitAlgebraic}}

Now let's consider numerical solutions of the model from the  preceding sections. We have applied to our neutrino models an adaptation of the  three distinct modified explicit methods that were developed and used to solve  astrophysical thermonuclear networks in Refs.\  \cite{guidJCP,guidAsy,guidQSS,guidPE,guid2016,broc2015,haid2016}:  

\begin{enumerate}

 \item 
the  {\em explicit asymptotic approximation} (Asy) \cite{guidJCP,guidAsy},

\item
{\em quasi-steady-state approximation} (QSS) \cite{guidJCP,guidQSS}, and

\item
 {\em partial equilibrium approximation} (PE) \cite{guidJCP,guidQSS},

\end{enumerate}

\noindent
where the partial equilibrium method will be used in conjunction with 
either the asymptotic method (Asy+PE approximation) or the quasi-steady-state 
method (QSS+PE approximation). In general these methods use algebraic 
constraints to remove the three distinct sources of stiffness appearing in 
generic kinetic equations that were documented in Ref.\ \cite{guidJCP}, thus 
stabilizing the integration for (much) larger timesteps than would be 
possible for traditional explicit methods.  For  these methods applied to the neutrino transport problem 
our results indicate that algebraically stabilized explicit methods can take 
timesteps comparable to those for implicit methods, but can compute each 
timestep more efficiently because no matrix inversions are required.  Thus, 
these methods hold considerable promise for faster and more efficient solution 
of neutrino transport equations in astrophysical explosions and mergers than are 
now possible. This paper introduces our neutrino evolution formalism and 
reports on results for the explicit asymptotic approximation, with our results 
 with this formalism using the QSS, Asy+PE, and QSS+PE approximations 
to be reported separately.

\subsection{The Explicit Asymptotic Approximation\label{Asy}}

The explicit asymptotic approximation (Asy) is described in depth for  thermonuclear networks in Refs.\ \cite{guidJCP,guidAsy}.  Since a thermonuclear  network and the neutrino evolution network being considered here are of similar  form, much of the formalism developed in  Ref.\ \cite{guidAsy} applies here also with appropriate modification, and we  shall highlight only the equations specific to the present neutrino formalism. Let's solve \eq{dNdt4} for the number density in bin $i$,
\begin{equation}
    \ndensity_{i} = \frac{1}{\tilde{\kappa}_{i}}\left (\eta_{i} 
    - \frac{d\ndensity_{i}}{dt} \right),
\end{equation}
and consider the asymptotic limit  $d\ndensity_{i}/dt  \rightarrow 0$ corresponding to $\eta_{i} \approx \tilde{\kappa}_{i}\ndensity_{i}$. Using a finite-difference approximation for the derivative at a particular  iteration gives
\begin{equation}
    \frac{d\ndensity_{i}^{n}}{dt} = \frac{\ndensity_{i}^{n+1} - 
    \ndensity_{i}^{n}}{\Delta t} - \frac{\Delta 
    t}{2}\frac{d^{2}\ndensity^{n}_{i}}{dt^{2}} + \ldots ,
\label{expdN1.0}
\end{equation}
where $\Delta t$ is the timestep and dots represent higher-order terms. This permits an approximate expression for the number density $\ndensity_{i}^{n + 1}$  at the $n+1$ step to be constructed from quantities known at step $n$,
\begin{align*} 
    \ndensity_{i}^{n+1} &= \frac{1}{\tilde{\kappa}_{i}}\left (\eta_{i} - 
    \frac{d\ndensity_{i}^{n}}{dt} \right) 
    \\
    &= 
    \frac{\eta_{i}}{\tilde{\kappa}_{i}}- 
    \frac{1}{\tilde{\kappa}_{i}}\frac{d\ndensity_{i}^{n}}{dt} 
    \\
    &= 
    \frac{\eta_{i}}{\tilde{\kappa}_{i}}- 
    \frac{1}{\tilde{\kappa}_{i}}\left(\frac{\ndensity_{i}^{n+1} - 
    \ndensity_{i}^{n}}{\Delta t}\right),
\end{align*}
where only the first term in the expansion on the right side of \eq{expdN1.0} has been retained in the last step.
This may be solved for $\ndensity_{i}^{n+1}$ to give
\begin{align} 
    \ndensity_{i}^{n+1} &= \frac{\ndensity_{i}^{n} + \Delta t \eta_{i}}{1 
    + \tilde{\kappa}_{i}\Delta t}
    \nonumber
    \\
   & = \ndensity_{i}^{n} + 
    \frac{\Delta t}{1 + \tilde{\kappa}_{i} \Delta t} ( \eta_{i} - 
    \tilde{\kappa}_{i}\ndensity_{i}^{n} )
    \nonumber
    \\
    &= \ndensity_{i}^{n} + 
    \frac{\Delta t C_{i}^{n}}{1 + \tilde{\kappa}_{i}\Delta t},
    \label{asymptotic}
\end{align}
where \eq{dNdt4} was used in the last step.
This is an explicit method because $\ndensity_{i}^{n+1}$ can be computed  directly from quantities already known at step $n$, so no  iterations with implied matrix inversions are required in the solution. It could also be viewed as a hybrid implicit--explicit or semi-implicit method because it is derived from \eq{expdN1.0}, which  has both $\ndensity_{i}^{n+1}$ and $\ndensity_{i}^{n}$ on the right hand side of the equation. 
In \cite{Laiu_2021}, \eq{asymptotic} was used to solve \eq{dNdt4} implicitly with fixed-point iteration, and can therefore be viewed as a single Picard step.  
However we choose to refer to this solution as an algebraically stabilized explicit method. 
We note that if $\ndensity_{i}^{n}\in[0,1]$ then $\eta_{i}/\tilde{\kappa}_{i}\in[0,1]$.  
Then is is easy to show that $\ndensity_{i}^{n+1}\in[0,1]$ for \emph{any} $\Delta t$, which implies unconditional stability.

\subsection{The Explicit Forward Euler Algorithm}

The simplest explicit method is the forward Euler algorithm, which when applied to \eq{dNdt4} takes the form 
\begin{equation}
    \ndensity_{i}^{n + 1} = \ndensity_{i}^{n} + \Delta t C_{i}^{n},
    \qquad
    C_{i}^{n} \equiv d\ndensity_{i}^{n}/dt  = \eta_{i} -  \tilde{\kappa}_{i}\ndensity_{i}^{n}.
    \label{forwardEuler}
\end{equation}
Generally the explicit forward Euler method will be stable for timesteps less  than a critical timestep $\taucrit$ that is proportional to the inverse of the fastest  rate in the system, $\taucrit = 2/\max(\tilde\kappa_i)$, while the validity of  the asymptotic approximation increases for timesteps larger than $\taucrit$.  Thus, we adopt a criterion that if the desired timestep in the numerical integration is less than $\taucrit$  the forward Euler method of \eq{forwardEuler} is used, since it will be stable, but if the desired  timestep is greater than or equal to $\taucrit$ the explicit asymptotic approximation of \eq{asymptotic} is used instead, since the forward Euler method would be unstable for $\Delta t \ge \taucrit$.

\subsection{\label{conserveParticleNumber} Conservation of Particle Number}

Physically, it is desirable that our methods conserve particle number to 
machine precision.  We investigate this by first considering particle number 
conservation for the forward Euler method, for which the total particle number 
$N(t)$ at time $t$ is given by
\begin{equation}
    N(t) = \int_{\mathbb{R}^{+}} \ndensity(\epsilon, t) 
    dV_{\epsilon} \approx \sum_{i = 1}^{N_{b}} \ndensity_{i}(t)\Delta 
    V_{i}^{\epsilon}.
\end{equation} 
Multiplying both sides of \eq{forwardEuler} by $\Delta  V_{i}^{\epsilon}$ and summing over all $i$ gives
\begin{equation}
    N^{n + 1} = N^{n} + \Delta t \sum_{i = 1}^{N_{b}} 
    C_{i}^{n}\Delta V_{i}^{\epsilon}.
    \label{Ntot}
\end{equation}
The second term in this expression vanishes because
\begin{align} 
    \sum_{i = 1}^{N_{b}} C_{i}^{n}\Delta V_{i}^{\epsilon} &= \sum_{i = 1}^{N_b} 
    \sum_{k = 1}^{N_{b}} \big[(1 - 
    \ndensity_{i})\rincomponent ik \ndensity_{k} 
    \nonumber
    \\
    &\qquad\qquad - (1 - 
    \ndensity_{k})\routcomponent ik \ndensity_{i} \big]\Delta 
    V_{i}^{\epsilon} \Delta V_{k}^{\epsilon} 
    \label{2ndterm}
    \\ 
    &= \sum_{i = 1}^{N_{b}} \sum_{k = 
    1}^{N_{b}}(\rincomponent ik - \routcomponent ki)(1 - 
    \ndensity_{i})\ndensity_{k} \Delta V_{i}^{\epsilon} \Delta V_{k}^{\epsilon} 
    = 0,
    \nonumber
\end{align}
which results from expanding  summations, gathering like terms   $(1 - \ndensity_{i})\ndensity_{k}$, and noting from \eq{symmScattKern}  that $\rincomponent ik = \routcomponent ki$, where  
$$
    \rinoutcomponent ik \equiv \rinout (\epsilon_{i}, \epsilon_{k}) .
$$ 
Thus  $N^{n + 1} = N^{n}$ and \textit{the forward  Euler algorithm conserves particle number} at each timestep.

Now apply the same analysis to the explicit asymptotic approximation.
The analog of \eq{Ntot} is 
\begin{equation}
    N^{n + 1} = N^{n} + \Delta t \sum_{i = 1}^{N_{b}} 
    \frac{C_{i}^{n}}{1 + \tilde{\kappa}_{i} \Delta t}\Delta V_{i}^{\epsilon},
\end{equation}
and the denominator may be expanded as
\begin{equation}
    \frac{1}{1 + \tilde{\kappa}_{i} \Delta t} = \sum_{m = 
    0}^{\infty} (-1)^{m}(\tilde{\kappa}_{i} \Delta t)^{m},
\end{equation}
to obtain 
\begin{align}
    N^{n + 1} &= N^{n} + \Delta t \sum_{i = 1}^{N_{b}} \sum_{m = 
    0}^{\infty} C_{i}^{n} \Delta V_{i}^{\epsilon} (-1)^{m}(\tilde{\kappa}_{i} \Delta 
    t)^{m}
    \notag 
    \\ 
    &= N^{n} + \Delta t \sum_{i = 1}^{N_{b}} 
    C_{i}^{n}\Delta V_{i}^{\epsilon}
    \nonumber
    \\
    &\quad+ \sum_{i = 1}^{N_{b}} \sum_{m = 1}^{\infty} 
    C_{i}^{n} \Delta V_{i}^{\epsilon} (-1)^{m}(\tilde{\kappa}_{i} )^{m}(\Delta t)^{m 
    + 1}
    .
\label{AsyParticleConserve}
\end{align}
When $\tilde{\kappa}_{i}\Delta t<1$, this agrees with \eq{Ntot} up to $\mathscr{O}(\Delta t^{2})$ since the second  term will vanish due to the symmetry of the kernels, as demonstrated in  \eq{2ndterm}. The problem is the third term of \eq{AsyParticleConserve}: the approximation  is  expected to be valid for $\tilde{\kappa}_{i} \Delta t \gg 1$, so the \textit{expansion  doesn't converge} and  {\em  the explicit asymptotic approximation does not guarantee conservation of particle number.}  
However, as the population density approaches equilibrium, $C_{i}^{n}\to0$ and the third term can be small even when $\tilde{\kappa}_{i} \Delta t \gg 1$.  
As for the  thermonuclear  networks investigated in Refs.\ \cite{guidJCP,guidAsy,guidQSS,guidPE}, we may  deal with this by regulating the timestep such that the integration yields  conservation of particle 
number  at an acceptable level for a given problem, when  employing the explicit asymptotic approximation. 

\subsection{\label{timestepper} Integration Timestep Control}

As we have shown above, our explicit algebraic algorithms have the disadvantage that they are not all guaranteed to conserve particle number, but the error in particle number is controlled by the integration timestep size.  Thus in the FENN algorithms for neutrino population evolution (and in the FERN algorithms for thermonuclear networks) timesteps are chosen such that  at each point in the calculation the particle number is conserved locally within a tolerance specified by user-defined parameters, with those parameters chosen such that integrated global error in particle number is conserved to acceptable tolerance for the entire calculation (with acceptable error tolerance dictated by the nature of the problem being integrated). 

For the first integration timestep $\Delta t_0$ a suitable guess is used (for example $\Delta t_0 = 0.01t_0$, where $t_0$ is the initial time) for a trial timestep, and at each timestep after the first we choose a trial timestep predicted by the previous timestep (as described below). For each integration timestep two calculations are performed, one for the full trial timestep and one for two successive timesteps, each equal to half the trial timestep.  We then accept provisionally the results of the two half timesteps and test for particle number conservation.  If the local tolerance conditions are satisfied, the timestep is accepted.  If the error is too large an iteration on the preceding algorithm is performed for successively smaller timesteps until the tolerance condition is satisfied and that timestep  is accepted. Conversely, if the error is judged too small for maximum efficiency with acceptable error in the initial comparison, the iteration increases the timestep accordingly.  The code has the capability to recalculate reaction rates at each step of this iteration, but in the present calculations we recalculate fluxes for each iteration but use the same reaction rates for all calculations within the timestep.  We then use comparison of errors for the two half timestep and full timestep approximations of the accepted solution to estimate a local derivative of error with respect to timestep size, and use that to project a trial timestep for the next integration step that is expected to have acceptable error. This procedure is repeated at each integration timestep of the calculation.

\section{Calculations and Comparisons}

We shall now present comparisons of numerical solutions using the explicit  asymptotic method and those from several standard numerical methods  appropriate for stiff systems. The models given in Table  \ref{tb:neutrinoModels} correspond to representative conditions in various  zones of a core-collapse supernova simulation. The standard test problem will  assume the neutrino particle densities to be distributed over 40 energy bins, and for each case we integrate the relaxation of an initial  strongly peaked gaussian distribution to the equilibrium Fermi--Dirac  distribution, as illustrated schematically in Fig.\  \ref{fig:neutrinoRelaxation}.
%
%
\doublefig
    {neutrinoRelaxation}
    {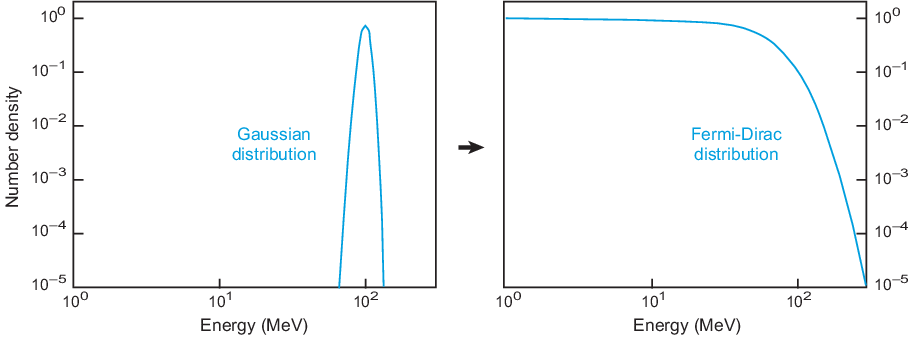}
    {0pt}
    {0pt}
    {1.00}
    {Test problem examined in this paper:  relaxation of an initial  gaussian neutrino energy distribution to a Fermi--Dirac  distribution through neutrino--electron scattering. Calculations were performed for the three temperature--density environments listed in Table \ref{tb:neutrinoModels}.
    }

As representative examples, timestepping and error plots (violation of  particle-number conservation) for the high-density Model I and low-density Model III from  Table \ref{tb:neutrinoModels} are displayed in Figs.\ \ref{fig:compareHighDens}  and \ref{fig:compareLowDens}.
%
%
\doublefig
    {compareHighDens}
    {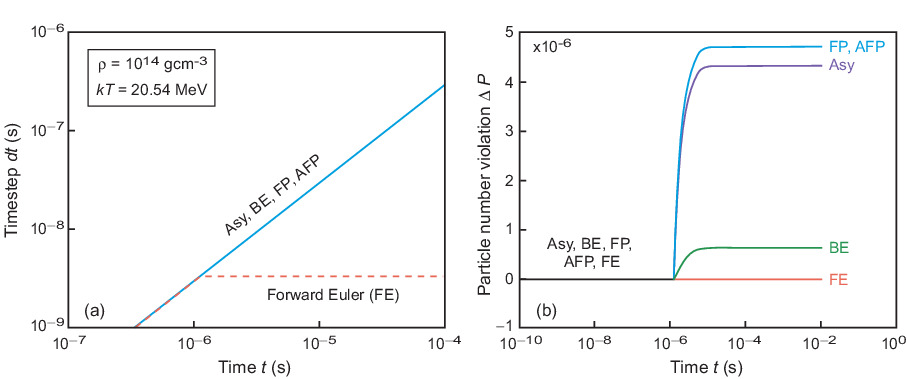}
    {0pt}
    {0pt}
    {1.0}
    {(a)~Integration timestep versus time for the high-density ($\rho=1\times  10^{14} {\rm\ g\,cm}^{-3}, \ kT = 20.54 {\rm\ MeV}$) Model I of Table  \ref{tb:neutrinoModels} using explicit asymptotic (Asy), backward Euler (BE),  fixed point (FP), accelerated fixed point (AFP), and forward Euler (FE) methods \cite{Laiu_2021}.   Parameters were chosen so that each of these methods (except for forward Euler) follows essentially the same timestepping curve $dt$ versus $t$.  Details concerning the  backward Euler, fixed point, and accelerated fixed point calculations may be  found in Ref.\ \cite{lac2020}. (b)~Violation of particle number conservation $\Delta P$ versus time for the  cases in (a) in units of $10^{-6}$. The timestepping curves in (a) for all  methods except forward Euler are virtually identical, and the final violation of particle number conservation in (b) ranges over negligible for forward Euler, $\sim 1\times  10^{-6}$ for backward Euler, and $\sim 4.5 \times 10^{-6}$ for the asymptotic,  fixed point, and accelerated fixed point methods. 
}
%
%
\doublefig
    {compareLowDens}
    {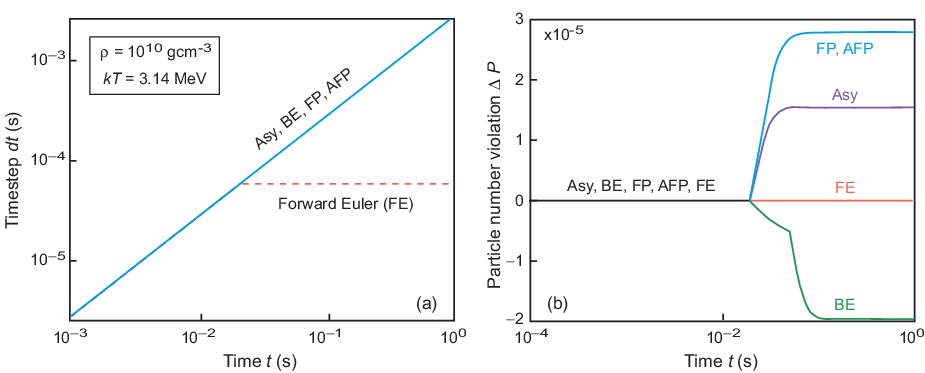}
    {0pt}
    {0pt}
    {1.00}
    {(a)~Integration timestep versus time for the low-density ($\rho=1\times 10^{10}  {\rm\ g\,cm}^{-3}, \ kT = 3.14 {\rm\ MeV}$) Model III of Table  \ref{tb:neutrinoModels} using explicit asymptotic (Asy), backward Euler (BE),  fixed point (FP), accelerated fixed point (AFP), and forward Euler (FE) methods.  Parameters were chosen so that each of these methods (except for forward Euler) follows essentially the same timestepping curve $dt$ versus $t$. Details concerning the backward Euler, fixed point, and accelerated fixed  point  calculations may be found in Ref.\ \cite{lac2020}. (b)~Violation of particle number conservation $\Delta P$ versus time for the  cases in (a) in units of $10^{-5}$. The timestepping curves in (a) for all  methods except forward Euler are virtually identical, and the final violation of particle number  conservation in (b) ranges over negligible for forward Euler to approximately $ -2\times  10^{-5}$ for backward Euler,  $1.5 \times 10^{-5}$ for asymptotic, and  $3 \times 10^{-5}$ for both fixed point and accelerated fixed point methods.
    }
Our results show in general that for evolution of neutrino energy distributions the explicit asymptotic method is capable of stable integration timesteps comparable in size to those of standard implicit methods, because the algebraic constraints remove substantial 
stiffness from the system of equations.  In particular, we find that

\begin{enumerate}

 \item 
 The algebraically stabilized explicit approach overcomes the disastrous  stability restriction that prevents increasing the timestep beyond a certain  time in the integration imposed generically by the stability constraint of  standard explicit methods; compare the forward Euler and explicit asymptotic  timestepping curves in Figs.\ \ref{fig:compareHighDens}(a) and  \ref{fig:compareLowDens}(a).

\item
When parameter choices yield similar timestepping for explicit asymptotic and  standard implicit methods, the errors in the calculated neutrino densities for  the asymptotic method are the same order of magnitude as the errors for standard  implicit methods; see the particle-number conservation error curves shown  in  Figs.\ \ref{fig:compareHighDens}(b) and  \ref{fig:compareLowDens}(b).

\end{enumerate}

\noindent
Explicit methods  can compute each timestep more efficiently than  implicit methods since no  matrix inversions are required.  This, coupled with the comparable timestepping  demonstrated in Figs.\ \ref{fig:compareHighDens}(a)  and \ref{fig:compareLowDens}(a), indicates that the explicit asymptotic method  for neutrino transport has the capability to out-perform standard implicit  methods, particularly for larger networks since the cost of matrix inversion increases rapidly with size of the matrices. This result is consistent  with those found previously for thermonuclear networks of similar structure as  the present neutrino transport network \cite{guidJCP,guidAsy,guidQSS,guidPE,fege2011,chup2008,brey2022,broc2015,  haid2016,guid2023a,brey2023}, but that were often larger and of even  greater stiffness.

\section{\label{speed-accuracyMeasures}Speed versus Accuracy for  Explicit Algebraic Methods}

Having established in some simple examples that the explicit asymptotic method shows promise for describing the evolution of neutrino populations, let us turn to a more detailed and quantitative comparison of explicit asymptotic simulations with traditional ones. In Fig.\ \ref{fig:timestepAndEnergyBins}
\singlefig
    {timestepAndEnergyBins}
    {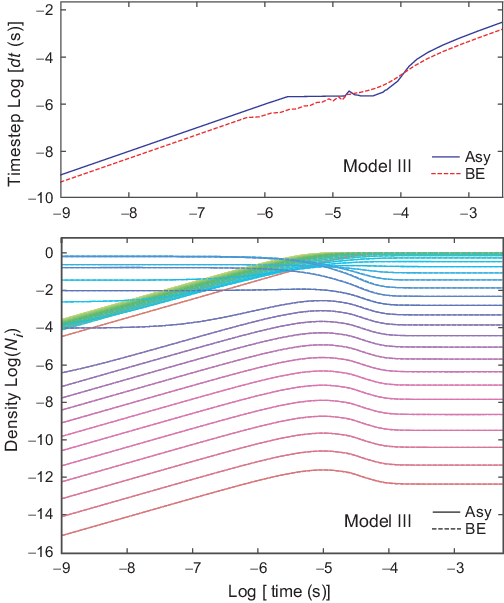}
    {0pt}
    {0pt}
    {1.0}
    {Neutrino number densities $\ndensity_i$ for 40 energy bins as a function of time $t$ given in seconds, calculated for the evolution of an initial gaussian distribution to a final Fermi--Dirac distribution at equilibrium as in Fig.\ \ref{fig:neutrinoRelaxation}, for Model I in Table \ref{tb:neutrinoModels}. (Lower Panel) Number density $\ndensity_i$  versus time $t$ in seconds.  Solid curves represent explicit asymptotic (Asy) results and dashed curves correspond to reference backward Euler (BE) implicit calculations.
    (Upper Panel) Timestepping for the calculation displayed in lower panel for the asymptotic approximation (Asy) and for the reference backward Euler calculation (BE).
    Calculations were performed with the FENN C++ codebase, except that reference backward Euler results were calculated using a standard integration routine in MatLab. 
    }
we illustrate the evolution of an initial gaussian energy distribution to a final Fermi--Dirac distribution (see Fig.\ \ref{fig:neutrinoRelaxation}) for the  temperature and density environment of Model I in Table \ref{tb:neutrinoModels}. We see that the explicit asymptotic calculation maintains competitive timestepping relative to a standard implicit backward Euler calculation across the entire calculation, while exhibiting rather good agreement with benchmark implicit calculations for the evolution of the neutrino densities.
Since explicit methods can compute each timestep faster than implicit methods, this suggests that these new explicit methods could be faster than standard implicit ones for computing the evolution of neutrino energy distributions in events such as core collapse supernova explosions and neutron star mergers. 

However, we have noted above that the explicit asymptotic method is an \textit{approximation} for which the amount of error introduced by the approximation can be \textit{controlled by placing restrictions on the timestepping}.  Thus we must address systematically whether the accuracy with competitive timestepping for explicit algebraic methods illustrated in Figs.\ $\ref{fig:compareHighDens}-\ref{fig:timestepAndEnergyBins}$ is sustainable over a range of realistic scenarios. With that in mind, let us turn to a systematic analysis of error versus speed in explicit asymptotic  evolution of neutrino populations.

\subsection{Controlled and Poorly Controlled Approximations}

As elaborated in Refs.\  \cite{guidJCP,guidAsy,guidQSS,guidPE,guid2016,broc2015,haid2016} and the present  paper, explicit algebraic methods are designed for use in large-scale  astrophysical simulations where typically a multi-dimensional radiation  hydrodynamics code is coupled in real time to neutrino transport and/or  thermonuclear burning kinetic networks. Often in such calculations evolving a realistic neutrino or thermonuclear network can dominate the computing budget, necessitating approximations.  In such an environment we desire that  approximations in the kinetic networks entail  uncertainties smaller  than the total errors and uncertainties from all other sources in the  calculation.  A reasonable criterion with  present technology is that errors of a few percent or less are acceptable for approximate  kinetic networks coupled to fluid dynamics. Here we  have given examples that explicit asymptotic methods can give errors less  than this threshold while exhibiting timestepping comparable to that   for standard implicit methods. For example, in Figs.\ \ref{fig:compareHighDens}  and \ref{fig:compareLowDens} the explicit asymptotic error in number-density conservation (a measure of  average error in  individual neutrino densities) is of order $10^{-5}$ to $10^{-6}$ for  timestepping equivalent to that of implicit methods, and  Fig.\ \ref{fig:timestepAndEnergyBins} will exhibit good agreement of explicit asymptotic and implicit methods with comparable timestepping.
This suggests that integration of  neutrino  kinetic equations can be  made more efficient by trading 

\begin{itemize}

\item
a {\em poorly controlled approximation,} common in current large-scale simulations (sufficient truncation and approximation of  realistic networks to allow them to be integrated quickly enough with standard methods), 

\item
for a  {\em controlled approximation} (algebraically stabilized explicit algorithms  applied to networks of \textit{realistic size and complexity}, with a quantitative error  
constraint dictated by the physics of the problem).

\end{itemize}

\noindent
 By such means, it should be  possible to  integrate many physically realistic problems that are presently intractable using standard algorithms on existing  machines.   This method of trading speed against accuracy for explicit algebraic approximations is documented in Ref.\ \cite{guid2023a} for thermonuclear networks.  We now investigate the tradeoff of speed for accuracy in explicit asymptotic neutrino simulations. Let's begin by defining quantitative measures of speed and accuracy.

\subsection{A Quantitative Measure of Accuracy}

 Explicit algebraic methods employ \textit{approximations} of the realistic problem, so the algorithm itself introduces some level of error. For example, we showed in  Section \ref{conserveParticleNumber} that the asymptotic approximation does not conserve particle number exactly. To examine the implications of employing an approximate solution one needs a systematic measure of the error associated with the approximation.
We may characterize the accuracy of an explicit algebraic evaluation of the neutrino evolution by integrating  deviation of the neutrino number densities  $\ndensity_i$ from benchmark results (which are taken to be exact), and computing the root mean square deviation in number densities summed over all species (energy bins) at that time.  This difference results in residuals 
$R_i(t)$ for each species $i$ in the network at every time $t$,
\begin{equation}
     R_i(t) \equiv |\ndensity_i(t) - \ndensity_i^0(t)|,
    \label{residuals}
\end{equation}
where $\ndensity_i(t)$ is the explicit asymptotic approximation for the number density in bin $i$ 
at time $t$ and $\ndensity_i^0(t)$ is the corresponding exact result. The 
root mean square (RMS) error $R(t)$ summed over all bins at time $t$ is then defined by  
\begin{equation}
    R(t) \equiv \sqrt{\sum_i R_i(t)} = \sqrt{\sum_i |\ndensity_i(t) - 
    \ndensity_i^0(t)|^2},
    \label{rms}
\end{equation}
where the sum is over all energy bins and $\sum_i \ndensity_i^0 = 1$. 
It will be shown later that the error $R(t)$ is typically finite only over a limited range of integrations times.
The total error per unit time $\error$ for integrating the network to equilibrium  is then approximated as
\begin{equation}
      \error= \frac{1}{\delta t} \sum_{j=j_0}^{j(t_{\rm eq})} R(t_j) \,\delta t_j,
\label{error-total}
\end{equation}
 where $t_j$ is the time at the $j$th plot output step,  $R(t_j)$ is given by \eq{rms}, $j_0$ is the first plot output step with finite $R(t)$, $j(t_{\rm eq})$ is the plot output step corresponding to the equilibration time, $\delta t_j = t_j - t_{j-1}$ is the time difference  between the $j$th and $(j-1)$th plot output steps,  and $\delta t \equiv t_{\rm eq} - t(j_0)$ is the total time between  onset of finite $R(t)$ and equilibration at $t_{\rm eq}$.
The error $\error$ computed from \eq{error-total} may be taken as a single number characterizing the global accuracy of a given calculation.

\subsection{A Quantitative Measure of Speed}

To evaluate the utility of explicit algebraic methods one needs a systematic measure of how long a calculation requires, relative to standard methods. 
The speed of a calculation can be parameterized conveniently in terms of the total number of integration steps required to complete a given simulation. This is an ``intrinsic'' measure, since it should be largely independent of computing platform.  The total integration time is then a function of the number of steps and the time required for each step.  Because standard implicit methods such as backward Euler typically take longer than an explicit method to compute each step, 

\begin{enumerate}

\item
if the algebraically stabilized explicit method requires a similar number or fewer integration steps than an implicit method, it will likely be faster, but  

\item
if it takes more integration steps than the implicit method, whether the  implicit or algebraic explicit method is faster will depend on the time to compute each timestep, which is platform dependent.

\end{enumerate}

\noindent
  The time to compute an explicit timestep relative to that for an implicit timestep on the same platform depends primarily on the additional linear algebra overhead incurred by the implicit method because of the required matrix inversions, which increases with network size. If the difference between explicit and implicit methods is approximated as being due entirely to the additional linear algebra overhead of the implicit method, then a speedup factor may be defined as $F\equiv 1/(1-f)$, where $f$ is the fraction of overall computing time spent by the implicit code in its linear algebra solver during a timestep.  Then we may expect that the explicit algebraic code can compute a single timestep approximately $F$ times faster than an implicit code in a given computational environment.
  
  The factor $F$ has been computed as a function of network size for thermonuclear networks in Ref.\ \cite{guidAsy}, using the implicit backward Euler code \textit{Xnet} \cite{hix1999} with a variety of linear solvers. The present neutrino network is similar in structure to a thermonuclear network, except that the neutrino network reaction matrix is dense but larger thermonuclear networks have quite sparse reaction matrices. However, the results for $F$ tabulated in Ref.\ \cite{guidAsy} in networks as small as 40 isotopes were determined using a dense-matrix solver in \textit{Xnet}. Thus we may expect that for the present 40-species neutrino network the value $F \sim 4$ inferred in thermonuclear networks is approximately valid and we shall assume in this discussion that the explicit asymptotic method can compute each timestep $\sim 4$ times faster than a corresponding implicit method on the same machine. Then the ratio of times required for integration of a given problem with explicit asymptotic and implicit backward Euler methods may be estimated as
  \begin{equation}
    \frac{\Delta t_{\rm EA}}{\Delta t_{\rm BE}} \simeq
    \frac{S_{\rm EA}}{F\cdot S_{\rm BE}} 
    \simeq 0.25 \, \left( \frac{S_{\rm EA}}{S_{\rm BE}} \right),
\label{wallClockRatio}
\end{equation}
where $\Delta t$ denotes elapsed wallclock time for the complete integration, $S$ is the number of integration steps, ``EA'' labels quantities from the explicit asymptotic calculation, ``BE'' labels quantities from the backward Euler calculation, and we have assumed a speedup factor $F\sim 4$ for the 40-species neutrino network employed in this work.

\section{\label{compareSpeedAccuracy}Trading Speed for Accuracy in Realistic Simulations}

Let us now use the tools developed in Section \ref{speed-accuracyMeasures} to evaluate speed versus accuracy for some explicit asymptotic (Asy) simulations of neutrino population evolution, using the simplified model described above that limits interactions to neutrino--electron scattering.  The benchmark that we shall employ for comparison  is an integration of the 40-species neutrino network using  a standard implicit backward Euler algorithm, under the same conditions as for the Asy calculation. 
Figure \ref{fig:neutrinoSpdAcc}
\doublefig
    {neutrinoSpdAcc}
    {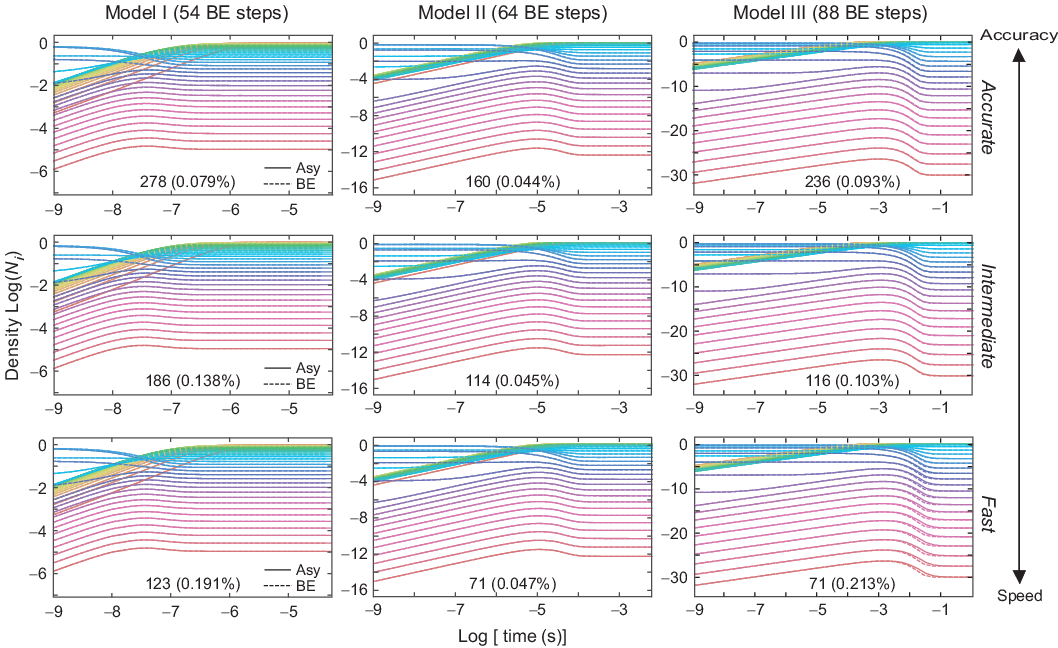}
    {0pt}
    {0pt}
    {1.00}
    {Speed versus accuracy using explicit algebraic methods for evolution of neutrino energy distributions.  In each plot 40 discrete energy bins for the neutrinos are tracked in time, assuming the only interaction to be neutrino--electron scattering.  Models I, II, and III are defined in Table \ref{tb:neutrinoModels}.  Solid curves are explicit asymptotic (Asy) calculations with timestepping parameters adjusted to emphasize speed (lower panels) versus accuracy (upper panels). This tradeoff can be adjusted continuously but for display purposes we have chosen three discrete parameter sets for each model, labeled ``Accurate'', ``Intermediate'', and ``Fast'', with Accurate being slower but more accurate, Fast being faster but less accurate, and Intermediate being intermediate in speed and accuracy. 
    Specifically, all parameters controlling the timestepper described in Section \ref{timestepper} were fixed for a given model, except that we varied the single parameter specifying the allowed deviation from particle number conservation in one timestep for  the asymptotic approximation to dial the ``Accurate'', ``Intermediate'', and ``Fast'' cases.   
    Dashed curves are reference calculations using an implicit backward Euler algorithm (labeled ``
    BE''). In most cases the solid and dashed curves are indistinguishable visually plotted at this scale,  with the exception of the Model III Fast Case, where small differences can be seen. The number of explicit asymptotic integration steps required is given as an integer at the bottom of each plot and that is followed in parentheses by the error calculated from \eq{error-total}, given as a percentage. 
    The number of integration steps required for the backward Euler (BE) reference integration was 54 steps for Model I, 64 steps for Model II, and 88 steps for Model III. All explicit Asymptotic calculations were performed with the FENN C++ codebase. Backward Euler reference curves were calculated with a standard integration package in MatLab.
    }
    illustrates the tradeoff of speed versus accuracy for some explicit algebraic calculations. In this figure the three columns of plots correspond to conditions associated with Models I, II, and III, respectively, in Table \ref{tb:neutrinoModels}, while the three successive rows of plots illustrate different choices of control parameters in the explicit asymptotic calculation that lead to decreasing speed but higher 
    accuracy for vertically successive plots. 
    
    Each plot is labeled near the bottom by the number of explicit asymptotic integration steps required, followed in parentheses by the RMS error per unit time $\error$  (expressed as a percentage) calculated from \eq{error-total} for the explicit algebraic calculation relative to the reference benchmark. 
    For example, in the third column of plots in Fig.\ \ref{fig:neutrinoSpdAcc} (corresponding to Model III from Table \ref{tb:neutrinoModels}), for successive plots from the top the number of explicit algebraic integration steps required decreases from 236 to 116 to 71,  while the error increases from 0.093\% to 0.103\% to 0.213\%.  Thus in this simple example we have gained a factor of more than three in speed at the expense of increasing the error by a little over a factor of two in dialing the integration parameters between ``Accurate'' and ``Fast'' sets, with the error in the ``Fast'' calculation still quite acceptable for coupling to a hydrodynamical simulation by our earlier arguments.
    
    The number of explicit asymptotic timesteps and the errors for all cases  in Fig.\ \ref{fig:neutrinoSpdAcc} are summarized in Table \ref{tb:stepsErrorComparison}. From Fig.\ \ref{fig:neutrinoSpdAcc} and Table \ref{tb:stepsErrorComparison}, the bottom row of plots in  Fig.\ \ref{fig:neutrinoSpdAcc} may be characterized  as faster but less accurate, the top row as slower but more accurate, and the middle row as intermediate in both speed and accuracy. 
We have exhibited the tradeoff of speed versus accuracy with three discrete examples in the rows of Fig.\ \ref{fig:neutrinoSpdAcc}, but  a continuum of possible parameter choices could be used to dial this tradeoff, as suggested by the double vertical arrow on the right side of Fig.\ \ref{fig:neutrinoSpdAcc}.
Let us now discuss in more depth the tradeoff of speed versus accuracy illustrated in  Fig.\ \ref{fig:neutrinoSpdAcc}.

\subsection{\label{accuracyAsy}Accuracy of the Explicit Algebraic Approximation}

The primary contribution to error in \eq{error-total} tends to occur in restricted ranges of integration time, with the error being negligible over much of the rest of the calculation.  For example, in Fig.\
 \ref{fig:RMSerror-neutrino}
\doublefig
{RMSerror-neutrino}
{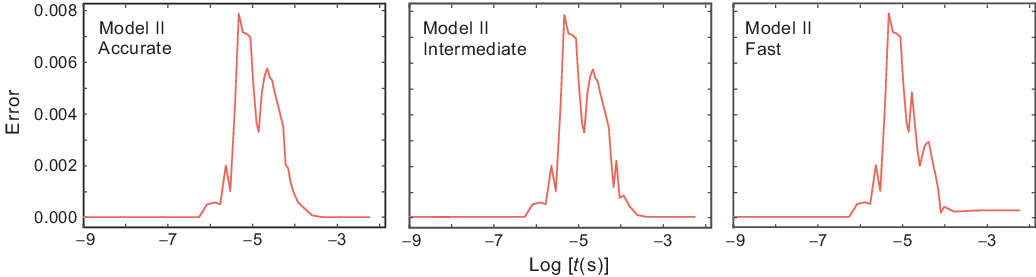}
{0pt}
{0pt}
{1.0}
{Root mean square error $ R(t)$ from \eq{rms} plotted versus time for the three Model II cases shown in Fig. \ref{fig:neutrinoSpdAcc}. For these examples almost all of the error accumulates between $\log t=-6$ and $\log t=-4$.}
we display the  RMS error $R(t)$ from \eq{rms} as a function of time for the Accurate, Intermediate, and Fast calculations of Model I that are displayed in Fig.\ \ref{fig:neutrinoSpdAcc}. In these examples we see that almost all of the error accumulates between $t \sim 10^{-6}$ and $t \sim 10^{-4}$ seconds.
In an  operator-split coupling to hydrodynamics the continuous integration  ranges used as illustration in Fig.\ \ref{fig:neutrinoSpdAcc} would be broken up into a  series of sequential piecewise integrations for a given fluid zone, with each piece representing integration of the network over one hydro timestep in the zone. Thus, significant error from the neutrino network for that zone will occur for only a restricted range of hydrodynamical  integration steps (those lying roughly between $10^{-6}$ and $10^{-4}$ seconds of elapsed fluid integration time in the preceding example).
We note that this could be used to optimize the tradeoff of speed versus  accuracy even further on a zone by zone basis in the coupled hydro--kinetic system.  We leave such optimizations for future applications of the basic ideas presented here to specific problems.

Of fundamental interest is a translation of neutrino network errors 
$\error$ defined in \eq{error-total} and displayed in  Fig.\ \ref{fig:neutrinoSpdAcc} and Table \ref{tb:stepsErrorComparison} into an overall error introduced into the coupled kinetic--hydrodynamical system by the explicit algebraic approximations.  The error computed from \eq{error-total} may be interpreted as the  RMS error per unit time, over the range of times for which $R(t)$ is finite. Therefore, no error is introduced for those hydro integration steps that do not overlap the regions of significant $R(t)$ in Fig.\
 \ref{fig:RMSerror-neutrino}. For those hydro timesteps that do overlap regions of significant $R(t)$, the error introduced by network approximations for a hydro timestep $\Delta t_{\rm hydro}$  will depend on the integrated value of $R(t)$ over the hydro timestep, with the hydro timestep depending on factors such as the method of hydro integration, zone sizes, and fluid characteristics at that time.  Thus it is case specific.  Nevertheless the relatively small values of $\error$ that can be chosen in 
Fig.\ \ref{fig:neutrinoSpdAcc} suggest that the maximum overall error can be controlled to a few percent or less while maintaining adequate speed, when this formalism is applied to specific problems.

\subsection{\label{speedAsy}Speed of the Explicit Algebraic Approximation}

The total time to integrate a problem is a function of the number of timesteps required and the time required to compute each timestep, with explicit methods generally computing a timestep faster than a corresponding  implicit method.
For each of the cases in Fig.\ \ref{fig:neutrinoSpdAcc}, we may use the number of integration steps required for the explicit asymptotic calculation and for a corresponding implicit backward Euler calculation in conjunction with \eq{wallClockRatio} to make a rough estimate of the overall relative speed of the explicit asymptotic and implicit calculations. Implicit backward Euler calculations required 54, 64, and 88 steps for integrating Models I, II, and III, respectively, over the time ranges in Fig.\ \ref{fig:neutrinoSpdAcc}, and the required asymptotic integration steps are given for each model and case in Fig.\ \ref{fig:neutrinoSpdAcc} and Table \ref{tb:stepsErrorComparison}.
\begin{table}[t]
    \centering
    \caption{Timesteps and errors $\error$ from \eq{error-total} for  Fig.\ \ref{fig:neutrinoSpdAcc}  \label{tb:stepsErrorComparison}} 
    \setlength{\tabcolsep}{4 pt}
    \begin{tabular}{lllllll}
    \hline
     & \ MODEL I & \  MODEL II & \ MODEL III  \\
     Case & Steps \ \ Error \ \  & Steps \ \ Error \ & Steps \ \ Error \\
    \hline
    Accurate & 278 \ \ \ \ \ 0.080\% & 160 \ \ \ 0.044\% & 236 \ \ \ 0.093\%\\
    Intermediate & 186 \ \ \ \ \ 0.138\% & 114 \ \ \ 0.045\% & 116 \ \ \ 0.103\%\\
    Fast & 123 \ \ \ \ \ 0.191\% & 71 \ \ \ \ \ 0.047\% & 71 \ \ \ \ \ 0.213\%\ \\
    \hline
    \end{tabular}
\end{table}
As an example, consider the rightmost column of plots in Fig.\ \ref{fig:neutrinoSpdAcc} corresponding to Model III, for which the backward Euler implicit calculation required 88 integration steps.  Then from \eq{wallClockRatio}, for the Accurate calculation (top row) the number of Asy integration steps required is 236 and the ratio of wallclock times for asymptotic integration relative to implicit integration on equivalent machines may be estimated as 
$$
\frac{\Delta t_{\rm EA}}{\Delta t_{\rm BE}} \simeq 0.25 \times \frac{236}{88} \simeq 0.67.
$$
Thus, for the Accurate calculation of Model III the explicit asymptotic integration is expected to be approximately $(0.67)^{-1} \simeq 1.5$ times as fast as the backward Euler calculation run on the same hardware. (The explicit method requires $\sim 2.7$ times as many integration steps, but computes each step $\sim4$ times faster than the implicit method.)  For the Intermediate Model III example a similar calculation gives
$$
\frac{\Delta t_{\rm EA}}{\Delta t_{\rm BE}} \simeq 0.25 \times \frac{116}{88} \simeq 0.33,
$$
so the Intermediate calculation is estimated to be about $(0.33)^{-1} \simeq 3$ times faster than the implicit BE calculation, while for the Fast Model III example
$$
\frac{\Delta t_{\rm EA}}{\Delta t_{\rm BE}} \simeq 0.25 \times \frac{71}{88} \simeq 0.20,
$$
and the Fast asymptotic calculation is estimated to be about $(0.20)^{-1} \simeq 5$ times faster than the implicit BE calculation.

Table \ref{tb:speedComparison}
\begin{table}[tbh]
    \centering
    \caption{Ratio of Asy to BE  speeds in Fig.\ \ref{fig:neutrinoSpdAcc} \label{tb:speedComparison}} 
    \setlength{\tabcolsep}{4 pt}
    \begin{tabular}{lccc}
    \hline
     Case & Model I & Model II & Model III  \\
    \hline
    Accurate     &    0.78 & 1.6  & 1.49     \\
    Intermediate     &    1.16  & 2.25  & 3.03     \\
    Fast     &    1.76 & 3.61  & 4.96      \\
    \hline
    \end{tabular}
\end{table}
summarizes the estimated overall speed of the explicit asymptotic (Asy) integration relative to backward Euler (BE) integration calculated from \eq{wallClockRatio} for all cases in Fig.\ \ref{fig:neutrinoSpdAcc}. On a single CPU the explicit asymptotic calculation is similar in speed to the implicit calculation for the ``Accurate'' calculations but is estimated to be 1-3 times faster than the implicit calculations for the ``Intermediate'' calculations, and faster than the implicit calculation by factors of approximately $~2-5$ for the various ``Fast'' calculations in Fig.\ \ref{fig:neutrinoSpdAcc}.

The numbers in Table \ref{tb:speedComparison} are  estimates but
these results illustrate  how explicit algebraic methods can be used to trade speed against the accuracy needed for a particular application in a controlled way. If one is coupling to a  hydrodynamical simulation of say a stellar explosion, then errors in the kinetic network of a few percent or less would likely be acceptable, given the errors and uncertainties associated with the whole system.  Thus, Table \ref{tb:speedComparison} indicates that the ``Intermediate'' and ``Fast'' calculations of Fig.\ \ref{fig:neutrinoSpdAcc} (middle and bottom row of plots) can compute our simple neutrino network $\sim 2-5$ times faster than an implicit code  (on a single CPU), with  error that is likely acceptable for coupling to hydrodynamics.  This advantage will grow for larger neutrino networks (more energy bins), since the Asy method requires no matrix inversions.  As we shall discuss in Section \ref{gpuacc}, this advantage   on a single CPU can increase dramatically if explicit algebraic approximations for kinetic networks are deployed in parallel on GPUs.

It should be noted that methods such as fixed-point or accelerated fixed-point that require no matrix inversions are faster than backward Euler methods for the problems addressed here \cite{Laiu_2021}, so they would likely compare more favorably than BE with our explicit algebraic calculations.  However, our purpose here is not a detailed comparison of methods but rather to demonstrate that explicit algebraic methods are generally competitive with currently used algorithms for evolution of neutrino populations, and could offer significant advantages, particularly when deployed on GPUs for large networks.

\section{\label{gpuacc}GPU Acceleration}

The present test cases for neutrino networks were implemented on single CPUs but we are porting our new explicit algebraic neutrino code to GPUs.  It has been shown in prior tests with thermonuclear networks that these explicit algebraic methods are extremely amenable to GPU acceleration because of their simplicity, which implies minimal memory footprint and communication overhead. For example, we have shown using an  NVIDEA K40 (Kepler) GPU that for a realistic 150-isotope thermonuclear network coupled through 1604 reactions under Type Ia supernova conditions,  $\sim 600$ independent networks could be deployed in parallel and  integrated on a single GPU in the same wall-clock time that a traditional  implicit code could integrate {\em one} such network \cite{guid2016,broc2015,haid2016}.  

The thermonuclear networks investigated in Refs.\ \cite{guidJCP,guidAsy,guidQSS,guidPE,fege2011,chup2008,brey2022,broc2015,  haid2016,guid2023a,brey2023} and the neutrino networks investigated here are structurally similar except that matrices associated with larger thermonuclear network are more sparse than those for the neutrino network.  Therefore, we expect that the explicit algebraic neutrino network is also likely to realize large efficiency gains when deployed on GPUs. Since the most powerful modern supercomputers acquire the bulk of their speed from  GPU accelerators, this bodes well for implementation of these new methods in future  petascale and exascale applications.

\section{Further Development}

The results presented here have offered proof of principle with a simple model that explicit algebraic methods may provide some advantages in describing the evolution of neutrino populations in stellar explosions and mergers.  As noted in the Introduction, this paper is the first step in developing and documenting the codebase FENN (\underline{F}ast \underline{E}xplicit \underline{N}eutrino \underline{N}etwork) for neutrino transport simulations in events like core collapse supernovae and binary neutron star mergers. Several additional steps are required to convert this proof of principle into a fully usable tool.

\begin{itemize}

\item
The neutrino couplings must be extended  beyond neutrino--electron scattering to include the full complement of expected neutrino--matter interactions.  

\item
This paper addresses only the explicit asymptotic (Asy) method of the explicit algebraic approximation.  In future work we will test the quasi-steady state (QSS) and the partial equilibrium (PE) methods also for neutrino networks.  Although in the examples discussed here the Asy approximation was adequate, prior experience with thermonuclear networks suggests that in more general situations the full set of Asy, QSS, and PE approximations may prove useful for the neutrino problem when the full range of neutrino--matter interactions is included.

\item
The examples presented here have used networks containing 40 ``species'' (neutrino energy bins). In future work we will investigate the scaling of execution time with number of species for the explicit algebraic neutrino network. This scaling has been established for thermonuclear networks in earlier work \cite{guidJCP,guidAsy}, but not yet for neutrino networks.  Because no matrix inversions are required, we may expect that the explicit algebraic methods will scale more favorably with network size than implicit methods, but we may expect some differences with the scaling found in thermonuclear networks because the matrices corresponding to large thermonuclear networks are more sparse than the matrices corresponding to the neutrino network.

\item
The FENN codebase will be ported to GPUs and optimized for the CPU--GPU architectures common on current petascale and exascale machines. We have laid the groundwork for that  in Refs.\ \cite{broc2015,haid2016} for explicit asymptotic thermonuclear networks using CUDA C code running on Nvidia GPUs, but that formalism must be adapted to neutrino networks and adapted to other GPUs.

\item
Once explicit algebraic GPU versions are in place for neutrinos, we intend to investigate the interface of FENN codes for neutrino networks (and FERN codes for thermonuclear networks \cite{guidJCP,guid2023a}) with multidimensional radiation hydrodynamics codes.  Many such codes are written primarily in FORTRAN (for example, Chimera \cite{brue2020}, FLASH \cite{fryx2000}, GenASiS \cite{budi2019}, and \texttt{thornado} \cite{pochik_etal_2021,Laiu_2021}), so this work will entail meshing the primarily C++ coding of FENN for neutrino networks and FERN for thermonuclear networks with the largely FORTRAN implementation of the hydrodynamics codes. 

\end{itemize}

\noindent
We then anticipate that these FENN and/or FERN tools may be used to simulate events like Type Ia (thermonuclear) supernovae, core collapse supernovae, binary neutron star mergers, nova explosions, and X-ray bursts, using neutrino and/or thermodynamics kinetic networks having physically realistic size and sophistication.

\section{Summary and Conclusions}

We have demonstrated for a representative neutrino transport problem that an  explicit asymptotic approximation leads to controlled errors that are of  acceptable size for coupling to large-scale hydrodynamics simulations, while displaying timestepping that is competitive with  standard implicit methods. Since explicit methods can compute each timestep  more efficiently than implicit methods because they do not require matrix  inversions, the explicit asymptotic method for neutrino transport shows promise  of being faster and more efficient than standard implicit methods in many large-scale simulations.  Furthermore, we may expect the relative advantage of  these new explicit methods to increase for larger networks because they avoid  matrix inversions, and for deployment on GPUs  because they entail relatively small memory and communication requirements.  Thus, we propose  that algebraically stabilized explicit methods may be capable  of integrating many realistic neutrino (and thermonuclear) networks coupled to  multidimensional hydrodynamics on existing high-performance platforms. 

With the proliferation of exascale machines and the development of algebraically stabilized explicit algorithms optimized for petascale and exascale architectures,  it  may be expected that even more realistic simulations  will become possible. To that end, we are developing  two primarily C++ codebases to exploit explicit algebraic methods:

\begin{itemize}

\item
FENN (\underline{F}ast \underline{E}xplicit \underline{N}eutrino \underline{N}etwork) for explicit algebraic solutions of astrophysical neutrino networks, described in this paper and Ref.\ \cite{lac2020}, and

\item
FERN (\underline{F}ast \underline{E}xplicit \underline{R}eaction \underline{N}etwork) for explicit algebraic solutions of astrophysical thermonuclear networks, described in Refs.\ \cite{guidJCP,guidAsy,guidQSS,guidPE,guid2023a}.

\end{itemize}

\noindent
These packages will be made available to the astrophysics community as documented open-source code for rapid solution of large kinetic networks coupled to fluid dynamics, and to the broader scientific community as an open-source template for problems in various disciplines that could profit from the scaling and efficiency gains inherent in algebraically stabilized explicit integration, particularly when they are deployed on modern GPUs.

\acknowledgments
We thank  Ashton DeRousse and Jacob Gouge for  help  with some calculations, and  acknowledge LightCone Interactive LLC for partial financial support of this  work. 
EE acknowledges support from the NSF Gravitational Physics Theory Program (NSF PHY 1806692 and 2110177).

\bibliography{referencesNetworksMWG}
\end{document}